\documentclass{JHEP3}
\usepackage{amssymb}

 \title{ Quasinormal Modes of Self-Dual Warped AdS$_\textbf{3}$ Black Hole
 in Topological Massive Gravity }

 \author{Ran Li\\
 Institute of Modern Physics,\\
 Chinese Academy of Sciences,\\
  Lanzhou, 730000, Gansu, China\\
 \email{liran05@lzu.cn}}

 \author{ Ji-Rong Ren\\
 Institute of Theoretical Physics,\\
  Lanzhou University,\\
   Lanzhou, 730000, Gansu, China\\
 \email{renjr@lzu.edu.cn}}

\date{\today}

 \abstract{We consider the
 various perturbations of
 self-dual warped AdS$_3$ black hole
 and obtain the exact expressions
 of quasinormal modes by imposing the
 vanishing Dirichlet boundary
 condition at asymptotic infinity.
 It is expected that the quasinormal modes
 agree with the poles of retarded Green's functions
 of the dual CFT.
 Our results provide a quantitative test
 of the warped AdS/CFT correspondence.}

 \newpage
 \begin{document}

 \section{Introduction}\label{sec-intro}

 Topological massive gravity (TMG) is described by the theory
 of three dimensional Einstein gravity
 with a gravitational Chern-Simons
 correction and the cosmological constant \cite{TMG1,TMG2}.
 The well-known spacelike warped AdS$_3$ black hole
 \cite{warpedads} (previously obtained in \cite{clement}),
 which is a vacuum solution of topological massive gravity,
 is conjectured to be dual to a two dimensional
 conformal field theory (CFT) with non-zero left
 and right central charges \cite{TMGcentralcharge}.

 Recently, a new class of solutions in
 three dimensional topological massive gravity
 named as self-dual warped AdS$_3$ black hole
 is proposed by Chen et al in \cite{chenselfdual}.
 The metric is given by
 \begin{eqnarray}
 ds^2=\frac{1}{\nu^2+3}\left[-\left(x-x_+\right)\left(x-x_-\right)d\tau^2
 +\frac{1}{\left(x-x_+\right)\left(x-x_-\right)}dx^2\right.&&\nonumber\\
 +\left.\frac{4\nu^2}{\nu^2+3}\left(\alpha d\theta+
 \frac{1}{2}\left(2x-x_+-x_-\right)d\tau\right)^2
 \right]&&,
 \end{eqnarray}
 where $x_+$ and $x_-$ are  the location of the
 outer and inner horizons respectively, and
 we have set $l=1$ for simplicity.

 The self-dual warped AdS$_3$ black hole,
 which is asymptotic to the warped AdS$_3$ spacetime,
 is locally equivalent to spacelike warped  AdS$_3$ spacetime.
 This solution is of the isometry group U(1)$\times$SL(2, R).
 It is shown in \cite{chenselfdual} that,
 under the consistent boundary condition,
 the U(1) isometry is enhanced to a Virasoro
 algebra with nonvanishing left central charge,
 while the SL(2, R) isometry becomes trivial
 with the vanishing right central charge,
 \begin{eqnarray}
 c_L=\frac{4\nu}{\nu^2+3}\;,\;\;\;c_R=0\;.
 \end{eqnarray}
 It is conjectured that
 the self-dual warped $AdS_3$ black hole
 is dual to a two dimensional
 chiral CFT,
 which suggests a novel example
 of warped AdS/CFT dual.

 The left and right temperatures
 of CFT can not be read directly
 from the coordinates transformation
 of locally identification to warped
 $AdS_3$ space.
 They can be defined
 with respect to the Frolov-Thorne vacuum \cite{frolov}.
 Considering the quantum field with
 eigenmodes of the asymptotic
 energy $\omega$ and angular momentum
 $k$, and assuming that the left
 and right charges $n_L$, $n_R$ are
 $k$ and $\omega$, the corresponding Boltzmann factor
 in terms of these variables is given by
 $e^{-\frac{\omega-k\Omega}{T_H}}=
 e^{-\frac{n_L}{T_L}-\frac{n_R}{T_R}}$,
 where the left and right temperatures are defined by
 \begin{eqnarray}
 T_L=\frac{\alpha}{2\pi}\;,\;\;\;
 T_R=\frac{x_+-x_-}{4\pi}\;.
 \end{eqnarray}

 In this paper, we want to investigate
 another interesting aspect of
 the self-dual warped AdS$_3$ black hole
 inspired by the
 AdS/CFT corresponding.
 It is expected that the quasinormal modes
 exactly agree with the location of
 the poles of retarded Green's functions
 of the dual CFT.
 In an remarkable paper \cite{BTZ}
 (see also \cite{BTZ1}),
 the quantitative agreement is confirmed by
 the analytical calculations of quasinormal
 modes of various perturbations for BTZ black hole.
 More recently, it is shown that
 the warped AdS$_3$ black hole preserves the same
 property as BTZ black hole \cite{chenjhep,chenplb}.
 One can refer to \cite{Horowitz,chan,wang,Cardoso} for
 numerical investigations about this aspect.
 We have studied the
 various perturbations of
 self-dual warped AdS$_3$ black hole
 and found that the wave equations
 can be exactly solved
 by the hypergeometric function.
 This observation allows us to
 analytically calculate the
 quasinormal modes for various perturbations.
 Comparing with the previously results
 reported in \cite{chenjhep},
 we show that the quasinormal modes
 are just of the forms predicted by dual CFT.
 The results may provide a quantitative test
 of the warped AdS/CFT correspondence.

 This paper is organized as follows.
 In the following three sections,
 we will calculate the quasinormal modes of scalar,
 vector and spinor perturbations respectively
 and compare them with the prediction
 of warped AdS/CFT dual.
 The last section is devoted to conclusion
 and discussion.

 \section{Quasinormal modes of scalar field perturbation}

 In this section, we will calculate the quasinormal
 modes of scalar field perturbation in the background of
 self-dual warped AdS$_3$ black hole.
 Let us consider
 the scalar field $\Phi$ with the mass $m$ in the background
 of self-dual warped AdS$_3$ black hole,
 where the wave equation is given by the
 Klein-Gordon equation
 \begin{eqnarray}
 \left(\frac{1}{\sqrt{-g}}\partial_\mu\left
 (\sqrt{-g}g^{\mu\nu}\partial_\nu\right)-m^2\right)\Phi=0\;.
 \end{eqnarray}

 Because the self-dual warped AdS$_3$ black hole
 possesses two killing vectors $\partial_\tau$
 and $\partial_\theta$,
 the scalar field wave function $\Phi(\tau, x, \theta)$
 can be expanded in eigenmodes as following
 \begin{eqnarray}
 \Phi(\tau, x, \theta)=e^{-i\omega\tau+ik\theta}R(x)\;,
 \end{eqnarray}
 where $\omega$ and $k$ are the energy and
 angular momentum of scalar field, respectively.
 Substituting this expression for scalar field perturbation
 into the Klein-Gordon equation,
 one can get the radial wave equation
 \begin{eqnarray}
 \left[\partial_x\left((x-x_+)(x-x_-)\partial_x\right)
 +\frac{\left(\omega+\frac{x_+-x_-}{2\alpha}k\right)^2}{(x-x_+)(x_+-x_-)}
 -\frac{\left(\omega-\frac{x_+-x_-}{2\alpha}k\right)^2}{(x-x_-)(x_+-x_-)}\right]R(x)&&
 \nonumber\\
 +\left(\frac{3(\nu^2-1)}{4\nu^2}\frac{k^2}{\alpha^2}-\frac{1}{\nu^2+3}m^2\right)
 R(x)=0\;.&&
 \end{eqnarray}

 This equation can be analytically solved by the
 hypergeometric function.
 By changing the variable to
 \begin{eqnarray}
 z=\frac{x-x_+}{x-x_-}\;,
 \end{eqnarray}
 the radial wave equation can be rewritten in the form
 of hypergeometric equation
 \begin{eqnarray}
 z(1-z)\frac{d^2 R(z)}{dz^2}+
 (1-z)\frac{dR(z)}{dz}+\left(
 \frac{A_s}{z}+B_s+\frac{C_s}{1-z}\right)R(z)=0\;,
 \end{eqnarray}
 where the parameters $A_s$, $B_s$ and $C_s$ are given by
 \begin{eqnarray}
 A_s&=&\left(\frac{k}{2\alpha}+\frac{\omega}{x_+-x_-}\right)^2
 \;,\nonumber\\
 B_s&=&-\left(\frac{k}{2\alpha}-\frac{\omega}{x_+-x_-}\right)^2
 \;,\nonumber\\
 C_s&=&\frac{3(\nu^2-1)}{4\nu^2}\frac{k^2}{\alpha^2}-\frac{1}{\nu^2+3}m^2\;.
 \end{eqnarray}

 According to the definition,
 the quasinormal modes of black hole
 must be purely ingoing at the horizon.
 So we are just interested in the solution
 with the ingoing boundary condition
 at the horizon.
 The solution of
 radial wave equation with the ingoing
 boundary condition
 is explicitly given by the hypergeometric
 function
 \begin{eqnarray}
 R(z)=z^{\alpha_s}(1-z)^{\beta_s}F(a_s,b_s,c_s,z)\;,
 \end{eqnarray}
 where
 \begin{eqnarray}
 \alpha_s=-i\sqrt{A_s}\;,\;\;\;
 \beta_s=\frac{1}{2}-\sqrt{\frac{1}{4}-C_s}\;,
 \end{eqnarray}
 and
 \begin{eqnarray}
 c_s&=&2\alpha_s+1\;,\nonumber\\
 a_s&=&\alpha_s+\beta_s+i\sqrt{-B_s}\;,\nonumber\\
 b_s&=&\alpha_s+\beta_s-i\sqrt{-B_s}\;.
 \end{eqnarray}

 Using the following transformation relation
 of hypergeometric function \cite{book}
 \begin{eqnarray}
 &&F(a,b,c;z)=\frac{\Gamma(c)\Gamma(c-a-b)}{\Gamma(c-a)\Gamma(c-b)}
 F(a,b,a+b-c+1;1-z)\nonumber\\
 &&\;\;\;\;+(1-z)^{c-a-b}
 \frac{\Gamma(c)\Gamma(a+b-c)}{\Gamma(a)\Gamma(b)}
 F(c-a,c-b,c-a-b+1;1-z)\;,
 \end{eqnarray}
 one can find the leading asymptotic behaviour
 $(z\rightarrow 1)$ of the solution
 \begin{eqnarray}
 R(z)\simeq z^{\alpha_s}(1-z)^{\beta_s}\frac{\Gamma(c_s)\Gamma(c_s-a_s-b_s)}
 {\Gamma(c_s-a_s)\Gamma(c_s-b_s)}\;.
 \end{eqnarray}

 Next, in order to find the quasinormal modes,
 one has to impose the boundary condition at asymptotic
 infinity. The condition that the flux
 vanishes at asymptotic infinity
 is just a perfect one. In this paper,
 we will use the equivalent
 Dirichlet condition that the field is
 vanishing at asymptotic infinity.
 By imposing the vanishing Dirichlet boundary
 condition at infinity, one can find the following
 relation
 \begin{eqnarray}
 c_s-a_s=-n\;,\;\;\textrm{or}\;\;c_s-b_s=-n\;,
 \end{eqnarray}
 which give the quasinormal modes of scalar perturbation
 \begin{eqnarray}
 k=-i(2\pi T_L)(n+h_L)\;,\nonumber\\
 \omega=-i(2\pi T_R)(n+h_R)\;,
 \end{eqnarray}
 with the left and right conformal weights
 of the operator dual to scalar fields
 \begin{eqnarray}
 h_L^s=h_R^s=
 \frac{1}{2}+\sqrt{\frac{1}{4}+
 \frac{3(1-\nu^2)}{4\nu^2}\frac{k^2}{\alpha^2}+\frac{1}{\nu^2+3}m^2}\;.
 \end{eqnarray}

 One can see that these modes coincide with the
 poles in the retarded Green's function
 obtained in \cite{chenselfdual}.
 So, for the scalar perturbation,
 our calculation indicates that the quasinormal
 modes of self-dual warped black hole are
 exactly predicted by the dual CFT.

 \section{Quasinormal modes of vector field perturbation}

 In this section, we calculate the
 quasinormal modes of vector perturbation.
 We consider the massive vector
 field described by the first order differential equation
 \begin{eqnarray}
 \epsilon_{\lambda}^{\;\;\alpha\beta}
 \partial_{\alpha}A_{\beta}=-mA_{\lambda}\;.
 \end{eqnarray}
 Assuming that the vector field takes the form
 \begin{eqnarray}
 A_{\mu}=e^{-i\omega\tau+ik\theta}\phi_{\mu}(x)\;,
 \end{eqnarray}
 one can derive the equations of motion
 \begin{eqnarray}
 -m\phi_x&=&\frac{g_{\tau\tau}}{\sqrt{-g}}
 (ik\phi_\tau+i\omega\phi_{\theta})\;,\\
 \frac{d\phi_\tau}{dx}&=&\frac{g_{xx}}{\sqrt{-g}}
 \left[(-\frac{\omega k}{m}+mg_{\tau\theta})\phi_\tau
 -(\frac{\omega^2}{m}+mg_{\tau\tau})\phi_{\theta}\right]\;,\\
 \frac{d\phi_\theta}{dx}&=&\frac{g_{xx}}{\sqrt{-g}}
 \left[(\frac{k^2}{m}+mg_{\theta\theta})\phi_\tau
 -(-\frac{\omega k}{m}+mg_{\tau\theta})\phi_{\theta}\right]\;.
  \end{eqnarray}

 From Eq.(3.4) and (3.5),
 one can get the following wave equation for $\phi_\theta$
 after changing the variables to $z$
 \begin{eqnarray}
 z(1-z)\frac{d^2\phi_\theta}{dz^2}
 +(1-z)\frac{d\phi_\theta}{dz}
 +\left(\frac{A_\nu}{z}+B_\nu+\frac{C_\nu}{1-z}
 \right)\phi_\theta=0\;,
 \end{eqnarray}
 with
 \begin{eqnarray}
 A_\nu&=&\left(\frac{k}{2\alpha}+\frac{\omega}{x_+-x_-}\right)^2
 \;,\nonumber\\
 B_\nu&=&-\left(\frac{k}{2\alpha}-\frac{\omega}{x_+-x_-}\right)^2
 \;,\nonumber\\
 C_\nu&=&\frac{3(\nu^2-1)}{4\nu^2}\frac{k^2}{\alpha^2}
 -\frac{m^2-2m\nu}{\nu^2+3}\;.
 \end{eqnarray}
 The solution with the ingoing boundary
 condition at the horizon is given by
 \begin{eqnarray}
 \phi_\theta=z^{\alpha_\nu}(1-z)^{\beta_\nu+1}
 F(a_\nu+1,b_\nu+1,c_\nu,z)\;,
 \end{eqnarray}
 where
 \begin{eqnarray}
 \alpha_\nu=-i\sqrt{A_\nu}\;,\;\;\;
 \beta_\nu=-\frac{1}{2}+\sqrt{\frac{1}{4}-C_\nu}\;,
 \end{eqnarray}
 and
 \begin{eqnarray}
 c_\nu&=&2\alpha_\nu+1\;,\nonumber\\
 a_\nu&=&\alpha_\nu+\beta_\nu+i\sqrt{-B_\nu}\;,\nonumber\\
 b_\nu&=&\alpha_\nu+\beta_\nu-i\sqrt{-B_\nu}\;.
 \end{eqnarray}

 From Eq.(3.5), one can find
 \begin{eqnarray}
 \phi_\tau=\bar{A}_\nu\phi_\theta+\bar{B}_\nu\frac{\phi_\theta}{1-z}
 +\bar{C}_\nu z\frac{d\phi_\theta}{dz}\;,
 \end{eqnarray}
 where
 \begin{eqnarray}
 \bar{A}_\nu&=&-\frac{2m^2\nu^2\alpha(x_+-x_-)+\omega k(\nu^2+3)^2}
 {4m^2\nu^2\alpha^2+k^2(\nu^2+3)^2}\;,\nonumber\\
 \bar{B}_\nu&=&\frac{4m^2\nu^2\alpha(x_+-x_-)}
 {4m^2\nu^2\alpha^2+k^2(\nu^2+3)^2}\;,\nonumber\\
 \bar{C}_\nu&=&\frac{2m\nu\alpha(\nu^2+3)(x_+-x_-)}
 {4m^2\nu^2\alpha^2+k^2(\nu^2+3)^2}\;.
 \end{eqnarray}

 Using the facts \cite{book}
 \begin{eqnarray}
 azF(a+1,b+1,c+1,z)=cF(a,b+1,c,z)-cF(a,b,c,z)\;,
 \end{eqnarray}
 and
 \begin{eqnarray}
 a(1-z)F(a+1,b,c,z)=(c-b)F(a,b-1,c,z)-(c-a-b)F(a,b,c,z)\;,
 \end{eqnarray}
 the solution can be explicitly given by
 \begin{eqnarray}
 \phi_\theta&=&a_\nu z^{\alpha_\nu}(1-z)^{\beta_\nu+1}
 F(a_\nu+1,b_\nu+1,c_\nu,z)\;,\nonumber\\
 \phi_t&=&z^{\alpha_\nu}(1-z)^{\beta_\nu}
 \left\{\left[\bar{A}_\nu+(\alpha_\nu+
 \beta_\nu-b_\nu)\bar{C}_\nu\right]
 (c_\nu-b_\nu-1)F(a_\nu,b_\nu,c_\nu,z)\right.\nonumber\\
 &&+\left[2\beta_\nu(\bar{A}_\nu+(\alpha_\nu+
 \beta_\nu-b_\nu)\bar{C}_\nu)+a_\nu(c_\nu-a_\nu-1)\right]
 F(a_\nu,b_\nu+1,c_\nu,z)\nonumber\\
 &&\left. +a_\nu(\bar{B}_\nu+\beta_\nu\bar{C}_\nu)
 F(a_\nu+1,b_\nu+1,c_\nu,z)\right\}\;.
 \end{eqnarray}

 One can also find the leading asymptotic behaviour
 $(z\rightarrow 1)$ of the this solution
 \begin{eqnarray}
 \phi_\theta&\simeq&z^{\alpha_\nu}(1-z)^{-\beta_\nu}
 \frac{\Gamma(c_\nu)\Gamma(a_\nu+b_\nu-c_\nu+2)}
 {\Gamma(a_\nu)\Gamma(b_\nu+1)}\;,\nonumber\\
 \phi_t&\simeq&z^{\alpha_\nu}(1-z)^{-\beta_\nu}
 \left\{\left[\bar{A}_\nu+(\alpha_\nu+
 \beta_\nu-b_\nu)\bar{C}_\nu\right]
 (c_\nu-b_\nu-1)(1-z)
 \frac{\Gamma(c_\nu)\Gamma(a_\nu+b_\nu-c_\nu)}{\Gamma(a_\nu)\Gamma(b_\nu)}
 \right.\nonumber\\
 &&+\left[2\beta_\nu(\bar{A}_\nu+(\alpha_\nu+
 \beta_\nu-b_\nu)\bar{C}_\nu)+a_\nu(c_\nu-a_\nu-1)\right]
  \frac{\Gamma(c_\nu)\Gamma(a_\nu+b_\nu-c_\nu+1)}{\Gamma(a_\nu)\Gamma(b_\nu+1)}
 \nonumber\\
 &&\left. +(\bar{B}_\nu+\beta_\nu\bar{C}_\nu)
 (1-z)^{-1}
 \frac{\Gamma(c_\nu)\Gamma(a_\nu+b_\nu-c_\nu+2)}{\Gamma(a_\nu)\Gamma(b_\nu+1)}
 \right\}\;.
 \end{eqnarray}
 By imposing the vanishing Dirichlet boundary
 condition at infinity, one can find the following
 relation
 \begin{eqnarray}
 a_\nu=-n\;,\;\;\textrm{or}\;\;b_\nu+1=-n\;,
 \end{eqnarray}
 which give the quasinormal modes of vector perturbation
 \begin{eqnarray}
 k&=&-i(2\pi T_L)(n+h_L^\nu)\;,\nonumber\\
 \omega&=&-i(2\pi T_R)(n+h_R^\nu)\;,
 \end{eqnarray}
 with the left and right conformal weights of the
 operator dual to vector field
 \begin{eqnarray}
 h_L^\nu=h_R^\nu+1=
 \frac{1}{2}+\sqrt{\frac{1}{4}+
 \frac{3(1-\nu^2)}{4\nu^2}\frac{k^2}{\alpha^2}+\frac{m^2-2m\nu}{\nu^2+3}}\;.
 \end{eqnarray}

 One can see that these modes are of the same form
 as the scalar case. But the conformal weight
 of operator dual to vector perturbation
 is slightly different from that obtained
 in \cite{chenjhep}.
 However, one can also conclude that,
 for the vector perturbation,
 the quasinormal modes are also
 predicted by the dual CFT.

 \section{Quasinormal modes of spinor field perturbation}

 In this section, we calculate the
 quasinormal modes of fermionic field perturbation
 in the background of self-dual warped AdS$_3$ black hole.
 We consider the spinor field $\Psi$
 with mass $m$, which obeys the covariant Dirac equation
 \begin{eqnarray}
 \gamma^a e^{\mu}_a\left(
 \partial_\mu+\frac{1}{2}\omega_{\mu}^{ab}\Sigma_{ab}
 \right)\Psi+m\Psi=0\;,
 \end{eqnarray}
 where $\omega_\mu^{ab}$ is the spin connection,
 which can be given in terms of the tetrad $e_a^\mu$,
 $\Sigma_{ab}=\frac{1}{4}[\gamma_a, \gamma_b]$, and
 $\gamma^0=i\sigma^2$, $\gamma^1=\sigma^1$,
 $\gamma^2=\sigma^3$,
 where the matrices $\sigma^k$ are the Pauli matrices.

 According to the metric of self-dual warped black hole,
 the tetrad field can be selected to be
 \begin{eqnarray}
 e^0&=&\sqrt{\frac{(x-x_+)(x-x_-)}{\nu^2+3}}\;d\tau\;,\nonumber\\
 e^1&=&\frac{1}{\sqrt{\nu^2+3}\sqrt{(x-x_+)(x-x_-)}}\;dx\;,
 \nonumber\\
 e^2&=&\frac{2\nu\alpha}{\nu^2+3}\;d\theta
 +\frac{\nu}{\nu^2+3}\left[
 (x-x_+)+(x-x_-)\right]d\tau\;.
 \end{eqnarray}
 By employing the Cartan structure equation
 $de^a+\omega^a_{\;\;b}\wedge e^b=0$,
 one can calculate the spin connection directly.
 The nonvanishing components of the spin connection
 are listed as follows
 \begin{eqnarray}
 \omega_\tau^{01}&=&\frac{3-\nu^2}{2(\nu^2+3)}\left(
 (x-x_+)+(x-x_-)\right)\;,\nonumber\\
 \omega_\tau^{12}&=&-\frac{\nu}{\sqrt{\nu^2+3}}
 \sqrt{(x-x_+)(x-x_-)}\;,\nonumber\\
 \omega_x^{02}&=&-\frac{\nu}{\sqrt{\nu^2+3}}
 \frac{1}{\sqrt{(x-x_+)(x-x_-)}}\;,\nonumber\\
 \omega_\theta^{01}&=&-\frac{2\nu^2\alpha}{\nu^2+3}\;.
 \end{eqnarray}
 The inverse of the tetrad field
 is given by
 \begin{eqnarray}
 e_0&=&\sqrt{\frac{\nu^2+3}{(x-x_+)(x-x_-)}}\left(
 \;\frac{\partial}{\partial\tau}
 -\frac{[(x-x_+)+(x-x_-)]}{2\alpha}
 \;\frac{\partial}{\partial\theta}
 \right)\;,\nonumber\\
 e_1&=&\sqrt{\nu^2+3}\sqrt{(x-x_+)(x-x_-)}
 \;\frac{\partial}{\partial x}
 \;, \nonumber\\
 e_2&=&\frac{\nu^2+3}{2\nu\alpha}
 \;\frac{\partial}{\partial\theta}\;.
 \end{eqnarray}

 Assuming that the spinor field takes the form
 $\Psi=(\psi_+(x),\psi_-(x))e^{-i\omega\tau+ik\theta}$
 and changing the variables to $z$,
 one can finally derive the following
 equations of motion after some algebra
 \begin{eqnarray}
 z^{\frac{1}{2}}(1-z)\frac{d\psi_+}{dz}+
 \left[\left(\frac{i\omega}{x_+-x_-}+\frac{ik}{2\alpha}
 +\frac{1}{4}\right)z^{-\frac{1}{2}}+
 \left(-\frac{i\omega}{x_+-x_-}+\frac{ik}{2\alpha}
 +\frac{1}{4} \right)z^{\frac{1}{2}}
 \right]\psi_+&&\nonumber\\
 +\left(-\frac{ik\sqrt{\nu^2+3}}{2\nu\alpha}
 -\frac{\nu}{2\sqrt{\nu^2+3}}+\frac{m}{\sqrt{\nu^2+3}}
 \right)\psi_-=0\;,&&\nonumber\\
  z^{\frac{1}{2}}(1-z)\frac{d\psi_-}{dz}+
 \left[\left(-\frac{i\omega}{x_+-x_-}-\frac{ik}{2\alpha}
 +\frac{1}{4}\right)z^{-\frac{1}{2}}+
 \left(\frac{i\omega}{x_+-x_-}-\frac{ik}{2\alpha}
 +\frac{1}{4} \right)z^{\frac{1}{2}}
 \right]\psi_-&&\nonumber\\
 +\left(\frac{ik\sqrt{\nu^2+3}}{2\nu\alpha}
 -\frac{\nu}{2\sqrt{\nu^2+3}}+\frac{m}{\sqrt{\nu^2+3}}
 \right)\psi_+=0\;.&&
 \end{eqnarray}
  The above equation can also be solved by the hypergeometric
  function. The solution with the ingoing boundary
  condition is explicitly given by
 \begin{eqnarray}
 \psi_+&=&c_fz^{\alpha_f+\frac{1}{2}}
 (1-z)^{\beta_f} F(a_f,b_f,c_f,z)\;,\nonumber\\
 \psi_-&=&z^{\alpha_f+\frac{1}{2}}
 (1-z)^{\beta_f}\left[c_f F(a_f,b_f,c_f,z)
 -b_f(1-z)F(a_f+1,b_f+1,c_f+1,z)\right]
 \nonumber\\
 &=&(c_f-b_f)z^{\alpha_f+\frac{1}{2}}
 (1-z)^{\beta_f} F(a_f+1,b_f,c_f+1,z)\;,
 \end{eqnarray}
 where
 \begin{eqnarray}
 \alpha_f&=&-i\left(\frac{\omega}{x_+-x_-}+\frac{k}{2\alpha}\right)
 -\frac{1}{4}\;,\nonumber\\
 \beta_f&=&\frac{1}{2}-
 \sqrt{\frac{3(1-\nu^2)}{4\nu^2}\frac{k^2}{\alpha^2}
 +\frac{m-\frac{\nu}{2}}{\nu^2+3}}\;,\nonumber\\
 \gamma_f&=&i\left(\frac{\omega}{x_+-x_-}-\frac{k}{2\alpha}\right)
 -\frac{1}{4}\;,\nonumber\\
 a_f&=&\alpha_f+\beta_f+\gamma_f\;,\nonumber\\
 b_f&=&\alpha_f+\beta_f-\gamma_f\;,\nonumber\\
 c_f&=&2\alpha_f+1\;.
 \end{eqnarray}

 One can also find the leading asymptotic behaviour
 $(z\rightarrow 1)$ of the this solution
 \begin{eqnarray}
 \psi_+&\simeq& c_f z^{\alpha_f+\frac{1}{2}}
 (1-z)^{\beta_f}\frac{\Gamma(c_f)\Gamma(c_f-a_f-b_f)}
 {\Gamma(c_f-a_f)\Gamma(c_f-b_f)}\;,\nonumber\\
 \psi_-&\simeq& z^{\alpha_f+\frac{1}{2}}
 (1-z)^{\beta_f}\frac{\Gamma(c_f+1)\Gamma(c_f-a_f-b_f)}
 {\Gamma(c_f-a_f)\Gamma(c_f-b_f)}\;.
 \end{eqnarray}
 By imposing the vanishing Dirichlet boundary
 condition at infinity, one can find the following
 relation
 \begin{eqnarray}
 c_f-a_f=-n\;,\;\;\;c_f-b_f=-n\;,
 \end{eqnarray}
 which give the quasinormal modes of spinor perturbation
 \begin{eqnarray}
 k&=&-i(2\pi T_L)(n+h_L^f)\;,\nonumber\\
 \omega&=&-i(2\pi T_R)(n+h_R^f)\;,
 \end{eqnarray}
 with the left and right conformal weight
 of the operator in the dual CFT dual to spinor fields
 \begin{eqnarray}
 h_L^f=h_R^f-\frac{1}{2}=
 \sqrt{\frac{3(1-\nu^2)}{4\nu^2}\frac{k^2}{\alpha^2}
 +\frac{m-\frac{\nu}{2}}{\nu^2+3}}\;.
 \end{eqnarray}

 Again, for the spinor perturbation,
 the quasinormal modes are of the same form
 as the scalar and vector cases.
 The conformal weight of operator in the
 dual CFT is also of the same form as that obtained
 in \cite{chenjhep}. It is shown that the quasinormal
 modes of spinor perturbation
 are also exactly predicted by the dual CFT.

 \section{Discussion}

 We have studied the scalar, vector and spinor field perturbations of
 self-dual warped AdS$_3$ black hole
 and obtained the exact expressions of
 the corresponding quasinormal modes.
 It is shown that the quasinormal modes
 of various perturbations
 are just of the forms predicted by the dual CFT.
 The results may provide a quantitative test
 of the warped AdS/CFT correspondence.

 At last, let us make some comments
 on the gravitational perturbations
 in warped AdS black hole background.
 TMG is an extension of three dimensional
 Einstein gravity with propagating degrees of freedom.
 It is interesting to consider gravitational
 perturbation and study the stability
 for black holes which is asymptotic to warped AdS$_3$.
 The differential equation of gravitational
 perturbation in TMG is generally third order,
 and is rather complicated to solve.
 The perturbation equation in the background of
 warped AdS black hole is given by
 \begin{eqnarray}
 \delta R_{\mu\nu}-\frac{1}{2}g_{\mu\nu}\delta R+2h_{\mu\nu}
 +\frac{1}{3\nu}\delta C_{\mu\nu}=0\;,
 \end{eqnarray}
 where
 \begin{eqnarray}
 \delta C_{\mu\nu}&=&\epsilon_{\mu}^{\;\;\alpha\beta}
 \nabla_{\alpha}\left(
 \delta R_{\beta\nu}-\frac{1}{4}g_{\beta\nu}\delta R
 -\frac{1}{4}Rh_{\beta\nu} \right)\nonumber\\
 &&-\epsilon_{\mu}^{\;\;\alpha\lambda}
 \delta\Gamma^{\beta}_{\alpha\nu}\left(
 R_{\lambda\beta}-\frac{1}{4}g_{\lambda\beta}R
 \right)\nonumber\\
 &&+\left(
 h_{\mu\gamma}\epsilon^{\gamma\alpha\lambda}
 -\frac{1}{2}h\epsilon_{\mu}^{\;\;\alpha\lambda}
 \right)\nabla_{\alpha}\left(
 R_{\lambda\nu}-\frac{1}{4}g_{\lambda\nu}R
 \right)\;,
 \end{eqnarray}
 and
 \begin{eqnarray}
 \delta\Gamma^{\lambda}_{\alpha\beta}&=&
 \frac{1}{2}\left(
 \nabla_{\beta}h^{\lambda}_{\;\;\alpha}
 +\nabla_{\alpha}h^{\lambda}_{\;\;\beta}
 -\nabla^{\lambda}h_{\alpha\beta}
 \right)\;,\nonumber\\
 \delta R_{\mu\nu}&=&
 \frac{1}{2}\left(
 \nabla_{\lambda}\nabla_{\mu}h^{\lambda}_{\;\;\nu}
 +\nabla_{\lambda}\nabla_{\nu}h^{\lambda}_{\;\;\mu}
 -\nabla^2 h_{\mu\nu}-\nabla_{\mu}\nabla_{\nu}h
 \right)\;,\nonumber\\
 \delta R&=&\nabla^{\mu}\nabla^{\nu}h_{\mu\nu}
 -\nabla^2 h\;.
 \end{eqnarray}
 Generally, it is believed that
 the equations of motion can be simplified
 after fixing the gauge conditions.
 This indeed happens in the case of BTZ black hole
 \cite{BTZTMG} and warped AdS$_3$ vacua
 \cite{stability}. For the case of black holes
 in warped AdS background, we do not
 know how to fix the gauge conditions and simplify the
 equations of motion.

 For BTZ black hole \cite{BTZTMG},
 which is asymptotic
 to AdS space, the equations of motion
 can be split into the second order
 equation for massless graviton
 and the first order equation for
 massive graviton.
 The quasinormal modes spectrum
 for gravitational perturbation can be
 constructed from the chiral
 highest weight modes by using the
 SL(2, R)$_L\times$SL(2, R)$_R$ symmetry of
 BTZ background. The extremal case
 have also been investigated in \cite{extremalbtz}.

 More recently, Chen et al in \cite{chenbinmode} have
 calculated the quasinormal modes of warped AdS
 black holes for scalar, vector and tensor
 perturbations by using the algebraic method developed
 in \cite{BTZTMG}. For the scalar and vector
 perturbations, our results for quasinormal modes are recovered.
 For the tensor perturbation,
 they considered the massive tensor field
 in warped AdS black holes background
 satisfying the following first order equation of motion
 \begin{eqnarray}
 \epsilon_{\mu}^{\;\;\alpha\beta}
 \nabla_{\alpha}h_{\beta\nu}+mh_{\mu\nu}=0\;,
 \end{eqnarray}
 and obtained the corresponding quasinormal modes.
 Their calculations depend on the observation
 of hidden conformal symmetry of warped AdS
 black holes in \cite{hcy,lihcy}.

 However, this formalism is hard to be generalized
 to calculate the gravitational perturbation
 in warped AdS black hole background
 because of the U(1)$_L\times$SL(2, R)$_R$ symmetry
 of the geometry of this class
 and the complexity of perturbation equation.
 So, for the warped AdS black holes,
 in order to calculate the quasinormal modes
 of gravitational perturbations,
 the more general method should be employed.
 It is definitely a hard work
 for future considerations.

 \section*{Acknowledgement}

 RL would like to thank Ming-Fan Li and Shi-Xiong Song for
 helpful discussions.
 The work of JRR is supported by the Cuiying Programme of Lanzhou
 University (225000-582404) and the Fundamental Research Fund for
 Physics and Mathematic of Lanzhou University(LZULL200911).


\begin{thebibliography}{99}

 \bibitem{TMG1}

  S. Deser, R. Jackiw and S. Templeton,
  \emph{Topologically massive gauge theories},
  Ann. Phys. \textbf{140}(1982)372.

 \bibitem{TMG2}

 S. Deser, R. Jackiw and S. Templeton,
 \emph{Three-Dimensional Massive Gauge Theories},
  Phys. Rev. Lett. \textbf{48}(1982)975.

 \bibitem{warpedads}

  D. Anninos, W. Li, M. Padi, W. Song, and A. Strominger,
  \emph{Warped $AdS_3$ Black Holes},
  JHEP, \textbf{0903}(2009)130.

   \bibitem{clement}

  K. A. Moussa, G. Clement and C. Leygnac, Class. Quantum Grav. \textbf{20}
(2003)L277;

  A. Bouchareb and G. Clement, Class. Quantum Grav. \textbf{24}(2007)5581;

  K. A. Moussa, G. Clement, H. Guennoune and C. Leygnac, Phys. Rev.
 D \textbf{78}(2008)064065.

 \bibitem{TMGcentralcharge}
 G. Compere and S. Detournay,
 \emph{Semi-classical Central Charge in Topologically Massive Gravity},
 Class. Quant. Grav. \textbf{26}(2009)012001;

 G. Compere and S. Detournay,
 \emph{Boundary Conditions for Spacelike and Timelike
 Warped $AdS_3$ Spaces in Topologically Massive Gravity},
 JHEP, \textbf{0908}(2009)092;

 M. Blagojevic and B. Cvetkovic,
 \emph{Asymptotic Structure of Topologically Massive
 Gravity in Spacelike Stretched AdS Sector},
 JHEP, \textbf{0909}(2009)006,



 \bibitem{chenselfdual}

 B. Chen, G. Moutsopoulos and B. Ning,
 \emph{Self-Dual Warped $AdS_3$ Black Holes},
 arXiv: 1005.4175 [hep-th].

 \bibitem{frolov}
  V. P. Frolov and K. S. Thorne,
 \emph{Renormalized Stress-Energy Tensor
 Near the Horizon of a Slowly Evolving,
 Rotating Black Hole}
 Phys. Rev. D \textbf{39}(1989)2125.

 \bibitem{BTZ}

 D. Birmingham, I. Sachs and S.N. Solodukhin,
 \emph{Conformal field theory interpretation of black hole
 quasi-normal modes},
 Phys. Rev. Lett. \textbf{88}(2002)151301.

 \bibitem{BTZ1}

 D. Birmingham,
 \emph{Choptuik scaling and quasinormal
 modes in the AdS/CFT correspondence},
 Phys. Rev. D \textbf{64}(2001)064024.


 \bibitem{chenjhep}

 B. Chen and Z.-B. Xu,
 \emph{Quasi-normal modes of warped black holes and
 warped AdS/CFT correspondence},
 JHEP, 0911(2009)091.

 \bibitem{chenplb}

  B. Chen and Z.-B. Xu,
 \emph{Quasinormal modes of warped AdS 3 black holes and AdS/CFT
 correspondence},
 arXiv:0901.3588.


 \bibitem{Horowitz}

 G. T. Horowitz and V. E. Hubeny,
 \emph{Quasinormal Modes of AdS Black Holes
 and the Approach to Thermal Equilibrium},
 Phys. Rev. D \textbf{62}(2000)024027.

 \bibitem{chan}

 S.F.J. Chan and R.B. Mann,
 \emph{Scalar Wave Falloff in Asymptotically Anti-de Sitter
 Backgrounds},
 Phys. Rev. D \textbf{55}(1997)7546.

 \bibitem{wang}

 B. Wang, C.-Y. Lin and E. Abdalla,
 \emph{Quasinormal modes of Reissner-Nordstr$\ddot{o}$m
 Anti-de Sitter Black Holes},
 Phys. Lett. B  \textbf{481}(2000)79.

 \bibitem{Cardoso}
 V. Cardoso and J. S. Lemos,
 \emph{Scalar, electromagnetic and
 Weyl perturbations of BTZ black holes: quasi normal modes},
 Phys. Rev. D \textbf{63}(2001)124015;
 V. Cardoso and J. S. Lemos,
 \emph{Quasi-Normal Modes of Schwarzschild
 Anti-De Sitter Black Holes: Electromagnetic
 and Gravitational Perturbations},
 Phys. Rev. D  \textbf{64}(2001)084017.

 \bibitem{book}

  M. Abramowitz and I.A. Stegun,
  \emph{Handbook of Mathematical Functions},
  Dover, New York, (1970).

 \bibitem{BTZTMG}

 I. Sachs and S. N. Solodukhin,
 \emph{Quasi-normal modes in topologically massive gravity},
 JHEP, 0808(2008)003.

 \bibitem{stability}

 D. Anninos, M. Esole and M. Guica,
 \emph{ Stability of warped $AdS_3$
 vacua of topologically massive gravity},
 JHEP, 0910(2009)083.

 \bibitem{extremalbtz}

 H. R. Afshar, M. Alishahiha and A. E. Mosaffa,
 \emph{Quasi-normal modes of extremal BTZ black holes
 in TMG},
 JHEP, 1008(2010)081.


 \bibitem{chenbinmode}

 B. Chen and J.Long,
 \emph{Hidden Conformal Symmetry and Quasi-normal Modes},
 arXiv:1009.1010[hep-th].


 \bibitem{hcy}

 R. Fareghbal,
 \emph{Hidden Conformal Symmetry of the Warped $AdS_3$ Black Hole},
 arXiv: 1006.4034[hep-th].



 \bibitem{lihcy}

  R. Li, M.-F. Li and J.-R Ren,
 \emph{Hidden Conformal Symmetry of Self-Dual
 Warped AdS$_3$ Black Holes in Topological Massive Gravity},
 arXiv:1007.1357[hep-th].




 \end{thebibliography}
 \end{document}